\documentclass[seceq]{ptptex}
\newcommand{\beq}{\begin{equation}}
\newcommand{\eeq}{\end{equation}}
\newcommand{\bea}{\begin{eqnarray}}
\newcommand{\eea}{\end{eqnarray}}

\renewcommand\Re{\operatorname{Re}}

\newcommand\mev{\,\mathrm{MeV}}
\newcommand\gev{\,\mathrm{GeV}}

\usepackage{graphicx}
\title{%        %You can use \\ for explicit line-break
Phases of QCD, Polyakov Loop and Quasiparticles\footnote{Work supported in part by BMBF, GSI and INFN}%
}

\author{%       %Use \scshape  for the family name
Wolfram \textsc{Weise}$^{1,}$\footnote{ e-mail address:
weise@ph.tum.de},\\  
Claudia \textsc{Ratti}$^{2}$ and Simon \textsc{R\"o{\ss}ner}$^{1}$}

\inst{%         %Affiliation, neglected when [addenda] or [errata]
$^1$ Physik-Department, Technische Universit\"at M\"unchen, \\
D-85747 Garching, Germany\\
$^{2}$ ECT*, I-38050 Villazzano (Trento), Italy
}

\abst{%         %this abstract is neglected when [addenda] or [errata]
QCD thermodynamics is studied using a model which combines Polyakov loop dynamics with 
spontaneous chiral symmetry breaking and its restoration (the PNJL model). The input is fixed 
entirely by pure-gauge lattice QCD results and by pion properties 
in vacuum. Successful comparisons with results from thermal lattice QCD are achieved,
including extrapolations to finite quark chemical potential. The phase diagram and selected susceptibilties for two quark flavors $(N_f = 2)$ are investigated with inclusion of diquark degrees of freedom.}

\begin{document}

\maketitle

\section{Introduction}

This presentation addresses the following principal question: is it possible to understand the results of lattice QCD (LQCD) thermodynamics in terms of properly selected quasiparticle degrees of freedom? A suitable framework for approaching this question is the PNJL model, a synthesis of Polyakov loop dynamics and the Nambu \& Jona-Lasinio (NJL) model. The Polyakov loop (or thermal Wilson line) is an indicator\cite{Pis00} of the confinement $\rightarrow$ deconfinement transition in QCD. The NJL model\cite{NJL61,NJL2}, on the other hand, provides a schematic description of the dynamical mechanism which drives spontaneous breaking of chiral symmetry in QCD. The PNJL model is thus based on two seemingly disconnected symmetries which characterise QCD in opposite limiting situations:\\

\begin{itemize}
\item{The $Z(3)$ center symmetry of the $SU(3)$ color gauge group is exact in the limit of pure gauge QCD, realised for {\it infinitely heavy} quarks. In the high-temperature, deconfinement phase of QCD this $Z(3)$ symmetry is spontaneously broken, with the Polyakov loop acting as the order parameter.}\\

\item{Chiral $SU(N_f)_R\times SU(N_f)_L$ symmetry is an exact symmetry of QCD with $N_f$ {\it massless} quark flavors.  In the low-temperature (hadronic) phase this symmetry is spontaneously broken down to the flavor group $SU(N_f)_V$ (the isospin group for $N_f = 2$ and the "eightfold way" for $N_f = 3$). As consequence there exist $2N_f + 1$ pseudoscalar Nambu-Goldstone bosons and the QCD vacuum is non-trivial. It hosts  quark condensates $\langle \bar{q} q \rangle$ which act as chiral order parameters.}\\ 
\end{itemize}

Confinement implies spontaneous chiral symmetry breaking, whereas the reverse is not necessarily true. Whether and under which conditions the chiral and deconfinement transitions coincide, as apparent in some lattice QCD results,  is thus a fundamental issue. 
Numerical simulations of QCD thermodynamics on the lattice\cite{Ka05,LP03} are the primary sources of information for our purpose. The equation of state of strongly interacting
matter is now at hand as a function of temperature $T$ and in a limited range of quark
chemical potentials $\mu$. Strategies for circumventing the fermion sign problem characteristic of finite $\mu$ are: improved multi-parameter re-weighting techniques\cite{FK02}, Taylor series expansion methods\cite{Allton02,Allton05,Ej06} and analytic continuation from imaginary chemical 
potential\cite{LP03,FP03,EL03}. LQCD data exist for the pressure, the energy and entropy densities, quark densities and selected susceptibilities.

The PNJL model is a useful device in order to interpret such LQCD results, to understand the underlying systematics in terms of quasiparticles and to extrapolate into regions not accessible by lattice computations.  This model can also contribute to the question about the entanglement between chiral and deconfinement transitions. It is designed much in analogy with a Ginsburg-Landau type approach: identifying the relevant order parameters as collective degrees of freedom which govern the dynamics and thermodynamics. Earlier versions of such a model have been discussed in refs.\cite{MO96,Fu04}. The present report summarises our recent results and developments of the two-flavor PNJL model\cite{RTW06,RRTW07,RRW07a,RRW07b} including successful comparisons with a variety of LQCD data. For related works see refs.\cite{GMMR06,MAS06,SFR06}.

\section{Sketch of the PNJL Model}

The two-flavor PNJL model\cite{RRW07a} is specified by the Euclidean action
\begin{equation} 
{\cal S}(\psi, \psi^\dagger, \phi)= \int _0^{\beta=1/T} d\tau\int_V d^3x \left[\psi^\dagger\,\partial_\tau\,\psi + {\cal H}(\psi, \psi^\dagger, \phi)\right] + \delta{\cal S}(\phi,T)
\label{eqn:action}
\end{equation} 
with the fermionic Hamiltonian density\footnote{$\vec{\alpha} = \gamma_0\,\vec{\gamma}$ and $\gamma_4 = i\gamma_0$ in terms of the standard Dirac $\gamma$ matrices.}:
\begin{equation}
{\cal H} = -i\psi^\dagger\,(\vec{\alpha}\cdot \vec{\nabla}+\gamma_4\,\hat{m}_0 -\phi)\,\psi + {\cal V}(\psi, \psi^\dagger)~,
\label{eqn:hamiltonian}
\end{equation}
where $\psi$ is the $N_f=2$ doublet quark field and $\hat{m}_0 = diag(m_u,m_d)$ is the quark mass matrix. 
The interaction ${\cal V}(\psi, \psi^\dagger)$, to be specified in detail later, includes four-fermion couplings acting in quark-antiquark and diquark channels. The quarks move in a background color gauge field $\phi \equiv A_4 = i A_0$, where $A_0 = \delta_{\mu 0}\,g{\cal A}^\mu_a\,t^a$ with the $\mathrm{SU}(3)_c$ gauge fields ${\cal A}^\mu_a$ and the generators $t^a = \lambda^a/2$. The matrix valued, constant field $\phi$ relates to the (traced) Polyakov loop as follows:
\begin{equation}
\Phi=\frac{1}{N_c}\mathrm{Tr}\left[\mathcal{P}\exp\left(i\int_{0}^{\beta}
d\tau A_4\right)\right]=\frac{1}{ 3}\mathrm{Tr}\,e^{i\phi}~,\label{eqn:polyakovloop}
\end{equation}
In a convenient gauge one can choose a diagonal representation, 
$\phi = \phi_3\,\lambda_3 +  \phi_8\,\lambda_8$, which leaves only $\phi_{3,8}$ as field variables representing the Polyakov loop.

\subsection{Polyakov loop effective potential}

The piece $\delta S = -\frac{V}{T}{\mathcal{ U}}(\phi,T)$ in (\ref{eqn:action}) carries information about the gluon dynamics which, in the present approach, is approximated by the dynamics of the Polyakov loop.  The effective potential $\mathcal{ U}$ models the region around the confinement-deconfinement transition in pure gauge QCD on the mean field, Ginsburg-Landau level. One can expect such an approach to work up to temperatures around twice $T_c$.  
At much higher temperatures a description of the thermodynamics entirely in terms of the Polyakov loop is no longer adequate as transverse gluons will become important.

\begin{figure}[htb]
\begin{minipage}[t]{68mm}
%\framebox[75mm]{\rule[-26mm]{0mm}{52mm}}
\includegraphics[width=7cm]{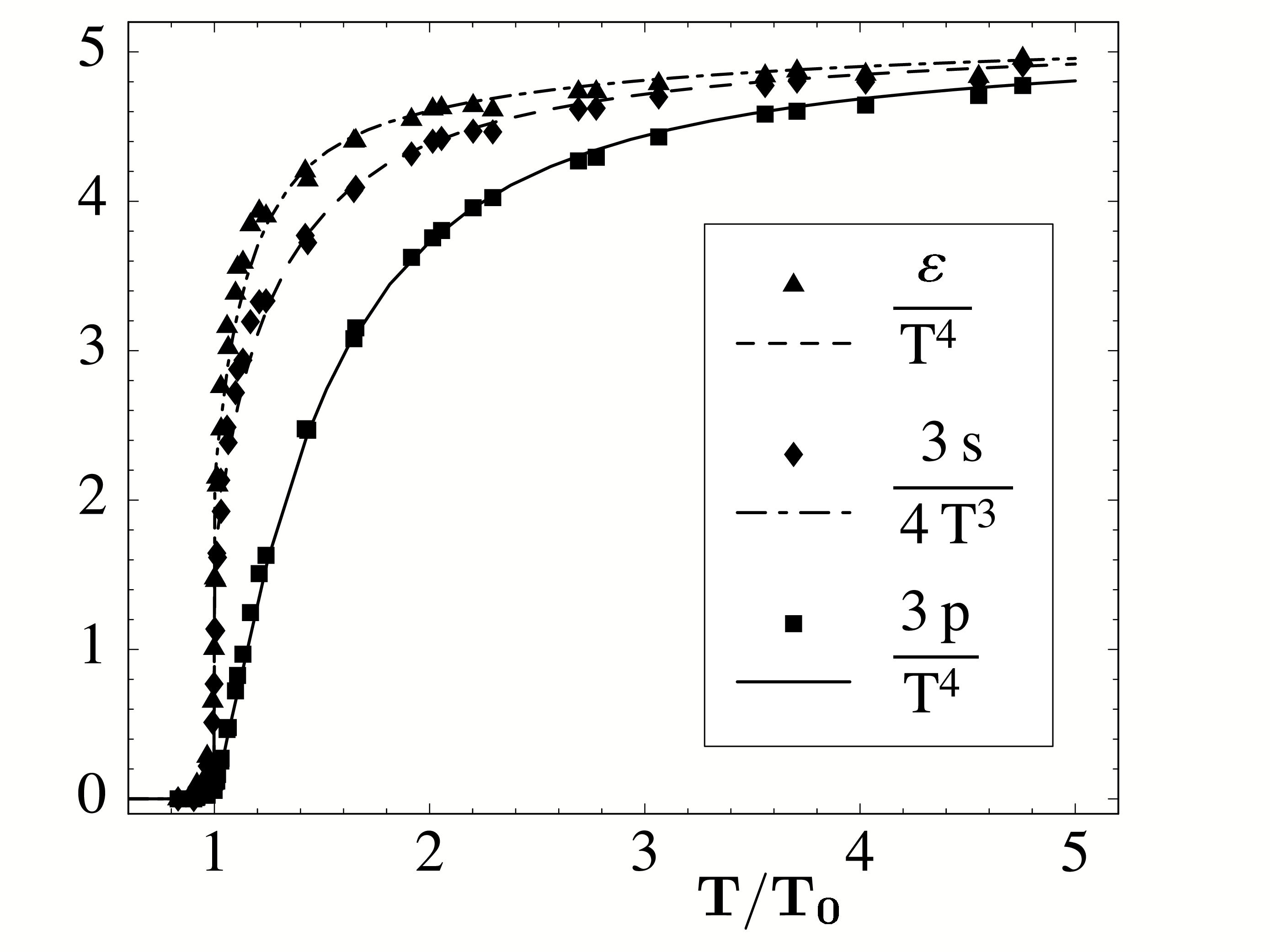}
\caption{Scaled energy density $\varepsilon$, entropy density $s$ and pressure $p$ as functions of temperature (given in units of the critical temperature $T_0 = 270$ MeV) of pure gauge QCD. Lattice results\cite{Boyd96} are compared with the Polyakov loop model (curves)\cite{RRW07a} with parameters as explained in the text.}
\label{fig:1}
\end{minipage}
\hspace{\fill}
\begin{minipage}[t]{68mm}
%\framebox[74mm]{\rule[-26mm]{0mm}{52mm}}
\includegraphics[width=7cm]{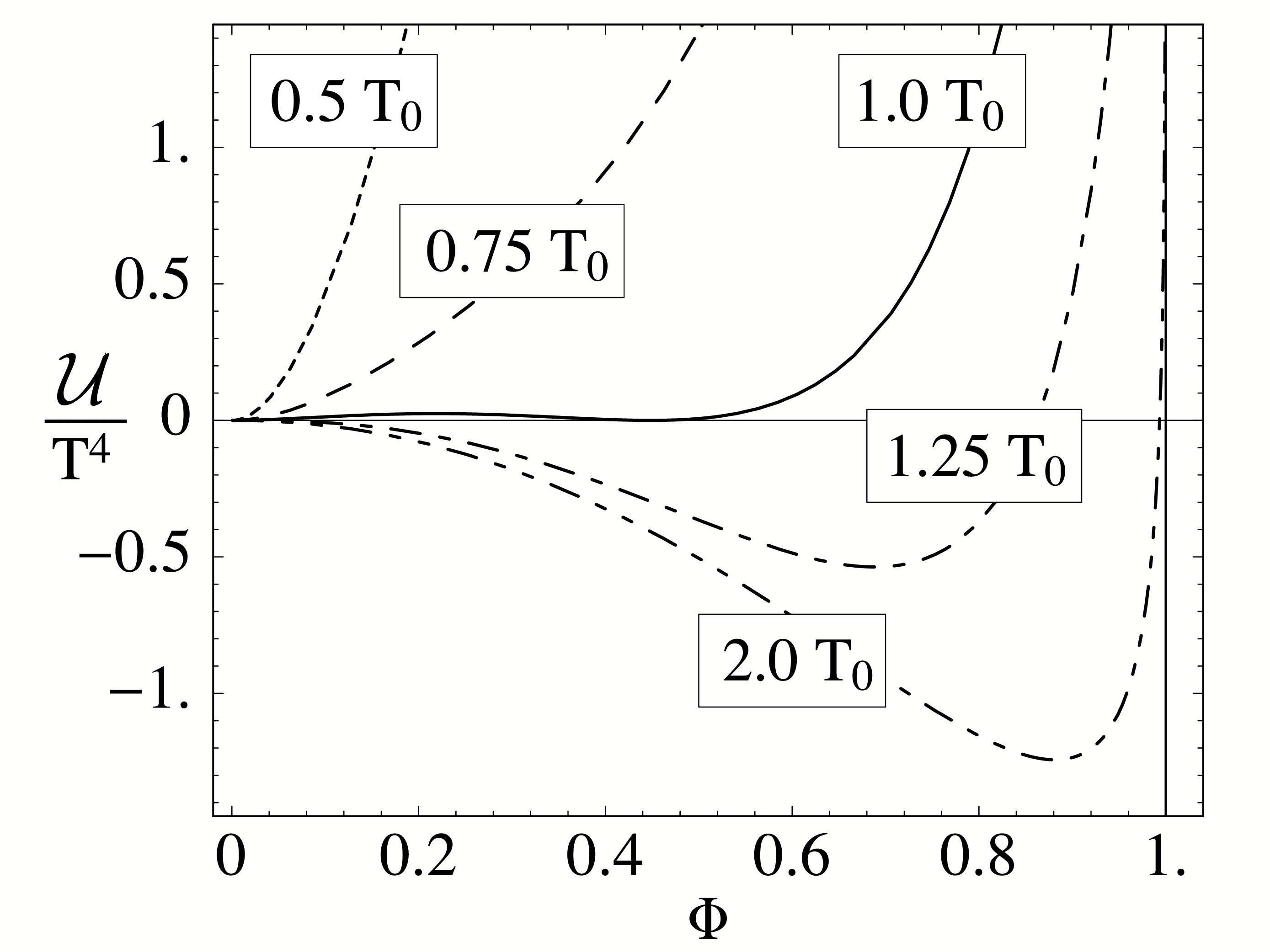}
\caption{Effective potential \ref{eqn:looppot} for the Polyakov loop $\Phi$ at different temperatures (given in units  of the critical temperature $T_0 = 270$ MeV) with parameters adjusted to reproduce the pure glue QCD equation of state (see Fig.\ref{fig:1}).}
\label{fig:2}
\end{minipage}
\end{figure}

The construction of the Landau effective potential describing Polyakov loop dynamics is guided by the
$\mathrm{Z}(3)$ center symmetry which transforms an element $u\in \mathrm{SU}(3)_c$ to $\exp(2\pi i n/3)u$ $(n = 1,2,3,...)$. The basic building blocks for such a potential are $\Phi^*\Phi$, $\Phi^3$ and ${\Phi^*}^3$ terms. In an earlier version of the model\cite{RTW06} a polynomial form for ${\cal U}$ was introduced and subsequently used also by other authors\cite{GMMR06}. However, this form does not satisfy the correct high-temperature limit. In the updated version\cite{RRTW07,RRW07a,RRW07b} the effective potential ${\cal U}$ has been designed to properly meet group theoretical constraints, with an ansatz motivated by the $\mathrm{SU}(3)$ Haar measure:
\begin{equation}
\frac{\mathcal{U}(\Phi,\,\Phi^*,\,T )}{T^4}=-\frac{1}{2}a(T)\,\Phi^*\Phi
+ b(T)\,\ln\left[1-6\,\Phi^*\Phi+4\left({\Phi^*}^3+\Phi^3\right)
-3\left(\Phi^*\Phi\right)^2\right]~~.
\label{eqn:looppot}
\end{equation}
The temperature dependent prefactors are written
\begin{align}
a(T) & = a_0+a_1\left(\frac{T_0}{T}\right)
+a_2\left(\frac{T_0}{T}\right)^2, & b(T)& =b_3\left(\frac{T_0}{T}
\right)^3.
\label{eqn:loopparam}
\end{align}
The logarithmic divergence in Eq.(\ref{eqn:looppot}) near $\Phi^*,\,\Phi\to 1$ makes sure that the Polyakov loop does not exceed constraints for the normalized trace of an $\mathrm{SU}(3)$ element.
The parameters are chosen such that the pressure $p = -\mathcal{U}(T)$ and related thermodynamical quantities reproduce the LQCD results for the pure glue equation of state \cite{Ka02} with its first order phase transition at a critical temperature $T_0\simeq 270$ MeV. The numerical values are given in ref.\cite{RRTW07,RRW07a} :
\begin{align}
a_0 &= 3.51\;, &a_1 &= -2.47\;, &a_2 &= 15.2\;, &b_3 &= -1.75\;.
\nonumber
\end{align}
Uncertainties in $a_{1,2,3}$ are estimated to be at the level of a few $\%$, while $a_0 =\frac{16\pi^2}{45}$ by virtue of the Stefan-Boltzmann limit. Figs.\ref{fig:1},\ref{fig:2} show the resulting fit to the pure gauge QCD equation of state and the Polyakov loop effective potential displaying the first order transition at $T = T_0$.
%figure-----------------------------------------------------------------------------------------------------------------------------
%figure-----------------------------------------------------------------------------------------------------------------------------

\subsection{NJL effective interaction}

The interaction $\mathcal{V}$ in Eq.(\ref{eqn:hamiltonian}) includes chiral $\mathrm{SU}(2)\times \mathrm{SU}(2)$ invariant four-point couplings of the quarks acting in quark-antiquark and diquark channels:
\begin{align}
\mathcal{V}= -\frac
{G}{2}\left[\left(\bar{\psi}\psi\right)^2+\left(\bar{\psi}\,i\gamma_5
\vec{\tau}\,\psi
\right)^2\right]
- \frac{H}{2}\left[\left(\bar{\psi}\,{\cal C}\gamma_5\tau_2\lambda_2
\,\bar{\psi}^{T}\right)\left(\psi^{T}\gamma_5\tau_2\lambda_2 {\cal C}
\,\psi\right)\right]~,
\label{eqn:V}
\end{align}
where $\mathcal{C}$ is the charge conjugation operator. One can think of Eq.(\ref{eqn:V}) as a subset in the chain of terms generated by Fierz-transforming a local color current-current interaction
between quarks, 
\begin{equation}
{\cal L}_{int} = - G_c(\bar{\psi}\gamma_\mu t^a\psi)(\bar{\psi}\gamma^\mu t^a\psi)~.
\nonumber
\end{equation}
In this case the coupling strengths in the quark-antiquark and diquark sectors are related by $G = \frac{4}{3}H$, the choice we adopt. The minimal ansatz (\ref{eqn:V}) for ${\cal V}$ is motivated by the fact that spontaneous chiral symmetry breaking is driven by the first term while the second term induces diquark condensation at sufficiently large chemical potential of the quarks. Additional pieces representing vector and axialvector $q\bar{q}$ excitations as well as color-octet diquark and  $q\bar{q}$ modes are omitted here. Their effects are not important in the present context.  

The NJL-model with $N_f = 2$ involves three parameters: the current quark mass $m_{u,d}$, a local four -quark coupling strength $G$ and a three-momentum cutoff $\Lambda$. As in refs.\cite{RTW06,RRTW07,RRW07a} we choose
\begin{align}
m_{u,d} \equiv m_0&= 5.5\mev\;,&G &= \frac{4}{3}H = 10.1\gev^{-2}\;,& \Lambda &= 0.65\gev\;.
\nonumber
\end{align}
These values were fixed to reproduce the pion mass and decay constant in vacuum, $m_\pi =$ 139.3 MeV, $f_\pi =$ 92.3 MeV, and the chiral condensate, $\langle\bar{\psi}_u\psi_u\rangle = - (251$ MeV)$^3$.

\section{Thermodynamics of the PNJL Model} 

In order to evaluate the thermodynamical properties of the PNJL model, new auxiliary fields  $(\sigma,\,\vec{\pi})$ and $(\Delta,\,\Delta^*)$ are introduced by bosonisation, absorbing the relevant quark-antiquark and diquark (antidiquark) correlations, respectively. The bosonised action involves the trace-log of the inverse Nambu-Gorkov matrix propagator which incorporates the couplings of quark quasiparticles, scalar and diquark fields. The pion field does not acquire an expectation value and the thermodynamic potential at mean field level becomes, after carrying out Matsubara frequency sums:
\begin{multline}
\Omega(T,\mu) =  \mathcal{U}\left(\Phi,\,\Phi^*,\,T\right)+\frac{\sigma^2}{2G}+
\frac{\Delta^*\Delta}{2H}\\
 - 2N_f\int\frac{d^3p}{\left(2\pi\right)^3}\sum_j \left\{
T\ln\left[1+e^{-E_j/T}\right]+\frac12 \Delta E_j
\right\}\;~.
\label{eqn:bosaction}
\end{multline}
The quasiparticle energies $E_j$  are determined analytically, with the results:
\begin{align}
E_{1,2}&=\varepsilon(\vec{p}\,)\mp\left(\mu+{2i\,\phi_8\over\sqrt{3}}\right),
&E_{3,...,6}=\sqrt{\left(\varepsilon(\vec{p}\,)\pm\left[\mu-\frac{i\,\phi_8}{\sqrt{3}}\right]\right)^2+|\Delta|^2}\,\mp i\,\phi_3~.
\nonumber
\end{align}
Here $\varepsilon(\vec{p}\,) = \sqrt{\vec{p}\,^2+m^2}$. The mass of the quark quasiparticles is generated dynamically by the gap equation, $m = m_0 + \sigma = m_0 -G\langle\bar{\psi}\psi\rangle$. The energy difference $\Delta E_j = E_j - \varepsilon_0 \pm \mu$ is taken with respect the free fermion energy, $\varepsilon_0 = \sqrt{\vec{p}\,^2+m_0^2}$.

At non-zero chemical potential $\mu$, the effective action is complex. The fermion sign problem persists in the PNJL model just as in QCD itself. As shown in ref.\cite{RRW07c} and following discussions in ref.\cite{DPZ05}, the consistent minimisation condition for the thermodynamic potential is
\begin{equation}
\frac{\partial\Re\Omega}{\partial \left(\sigma, \Delta, \phi_3, \phi_8\right)}=0~.
\label{eqn:mfeqn}
\end{equation}
In general, the Polyakov loop fields $\Phi$ and $\Phi^*$ (and their thermal expectation values) are different at non-zero quark chemical potential. This is primarily an effect of fluctuations of the fields around their mean values. The mean field approximation identifies the fields with their expectation values and implies\cite{RRW07c} $\Phi = \Phi^*$, or equivalently, $\phi_8$ = 0. 

\section{Results}

We now proceed to discuss selected examples and applications of PNJL thermodynamics, mostly in comparison with thermal LQCD results. 

\subsection{Chiral condensate and Polyakov loop}

\begin{figure}[htb]
\begin{minipage}[t]{65mm}
%\framebox[75mm]{\rule[-26mm]{0mm}{52mm}}
\includegraphics[width=6.4cm]{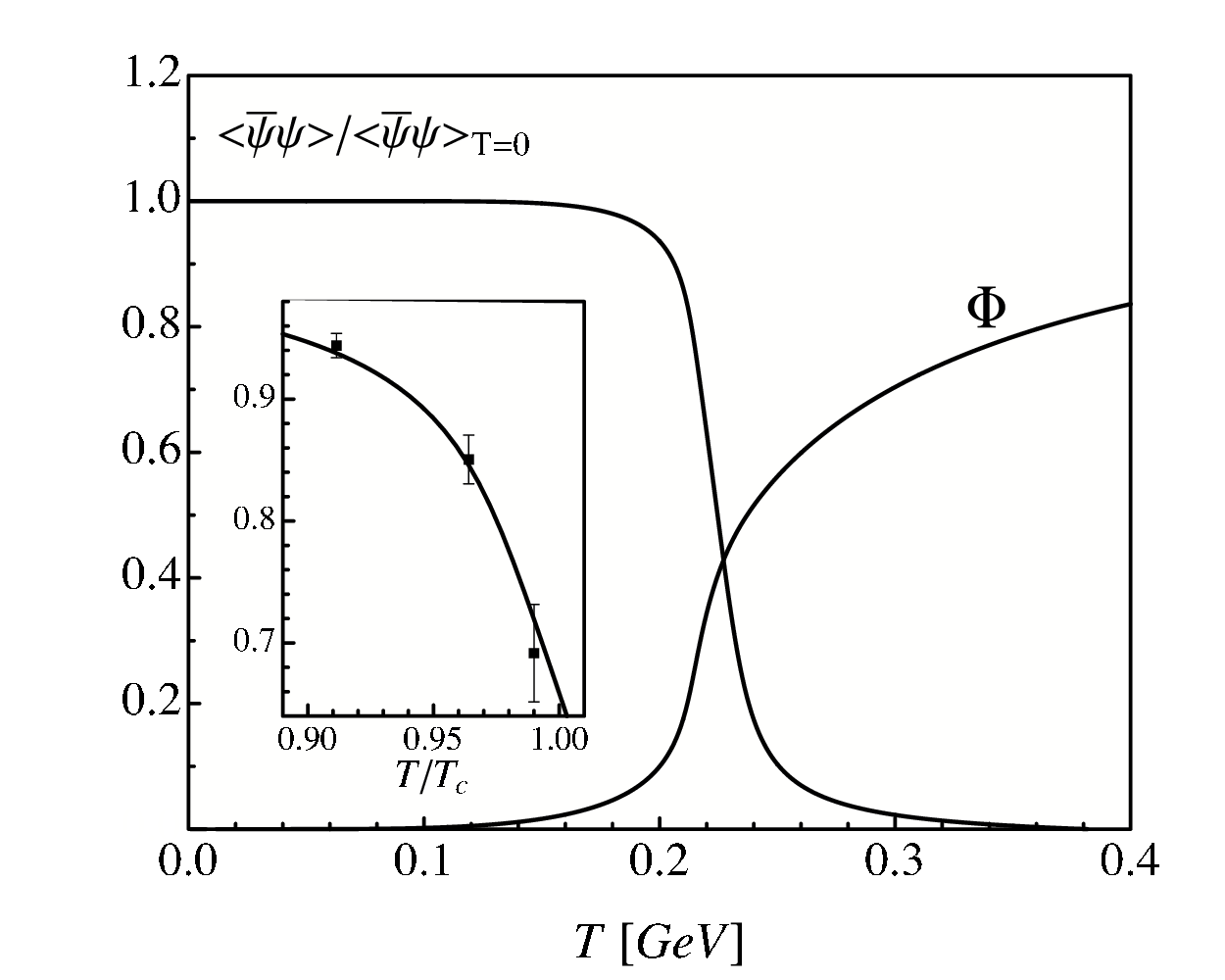}
\caption{Chiral (quark) condensate $\langle\bar{\psi}\psi\rangle$ and Polyakov loop $\Phi$ as functions of temperature at zero quark chemical potential, calculated using the PNJL model\cite{RTW06,RRTW07}. The inlay shows a comparison of the quark condensate with $N_f = 2$ lattice data taken from ref.\cite{Boyd95} .} 
\label{fig:3}
\end{minipage}
\hspace{\fill}
\begin{minipage}[t]{70mm}
%\framebox[74mm]{\rule[-26mm]{0mm}{52mm}}
\includegraphics[width=7.5cm]{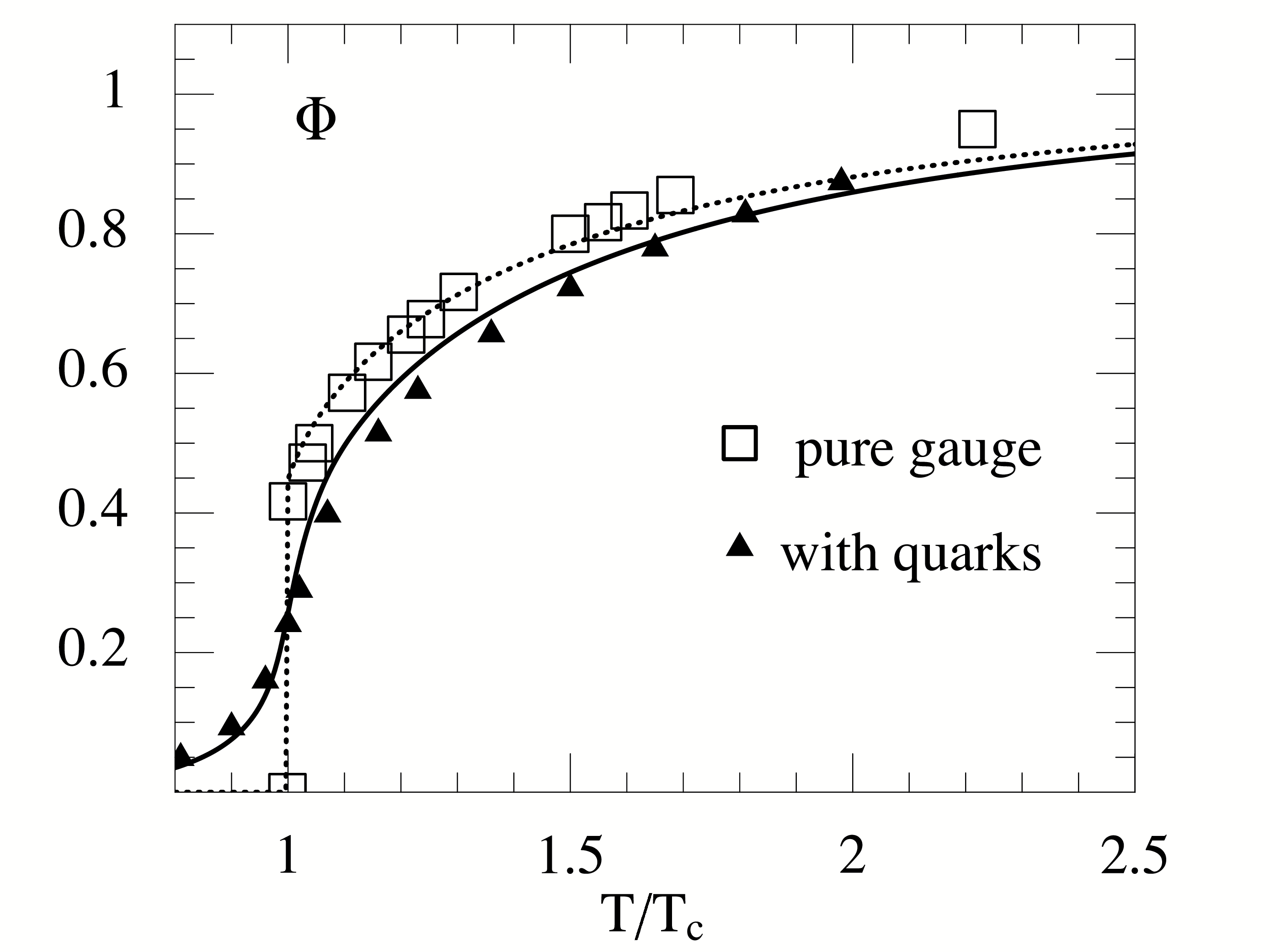}
\caption{Calculated temperature dependence of the Polyakov loop $\Phi$ in the pure gauge sector (dotted curve) using the effective potential of Fig.\ref{fig:2}, and prediction using the full PNJL model (solid curve)\cite{RRW07a} at zero quark chemical potential. Lattice data taken from refs.\cite{Ka02,KZ05} are shown for comparison.}
\label{fig:4}
\end{minipage}
\end{figure}

Consider first the chiral condensate $\langle\bar{\psi}\psi\rangle$ and the Polyakov loop $\Phi$ and their temperature dependence at zero quark chemical potential (see Fig.\ref{fig:3}). The original first-order deconfinement transition of pure gauge QCD is now turned into a crossover (see Fig.\ref{fig:4}) when quarks are introduced. The critical temperature is reduced from $T_0 \simeq 270$ MeV for the pure gauge system to a transition temperature $T_c \simeq 215$ MeV which almost coincides with the characteristic temperature for the crossover to chiral symmetry restoration. In the chiral limit with massless quarks $(m_0\rightarrow 0)$, the quark condensate vanishes at $T_c$ as a second order phase transition.  The crossover temperature found in the PNJL model is not far from recent lattice QCD results which give $T_c\simeq 200$ MeV for two flavors\cite{Ch06} and $T_c=192\pm 11$ MeV for $2+1$ flavors\cite{KZ05} .
%figure-----------------------------------------------------------------------------------------------------------------------------
%figure-----------------------------------------------------------------------------------------------------------------------------

\subsection{Finite quark chemical potential}

\begin{figure}[htb]
\begin{minipage}[t]{68mm}
%\framebox[75mm]{\rule[-26mm]{0mm}{52mm}}
\includegraphics[width=7cm]{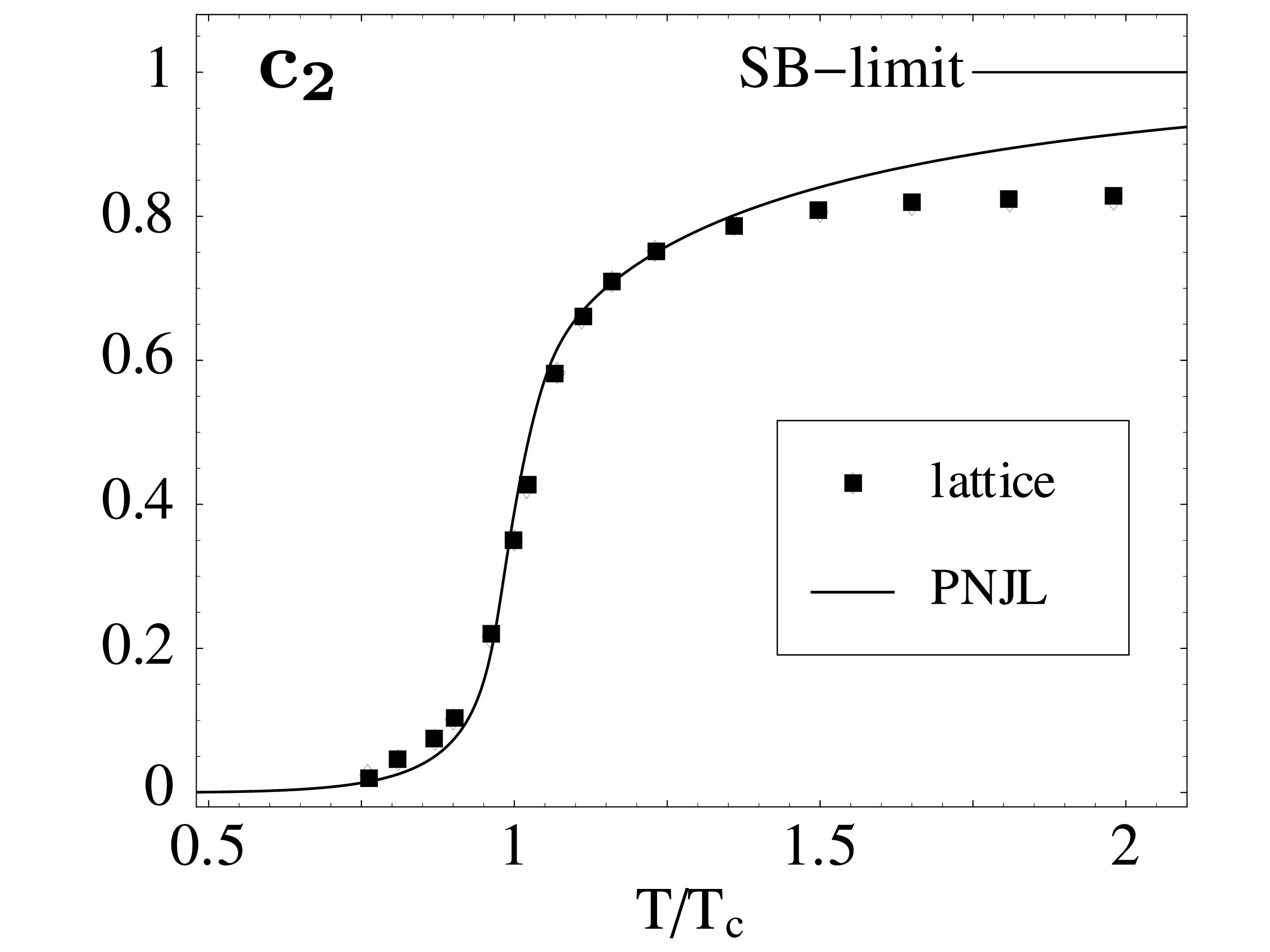}
\caption{Expansion coefficient $c_2(T)$ of Eq.(\ref{pressure}) predicted in the PNJL model\cite{RRTW07,RRW07a} and compared to lattice QCD results taken from ref.\cite{Allton05}.}
\label{fig:5}
\end{minipage}
\hspace{\fill}
\begin{minipage}[t]{68mm}
%\framebox[74mm]{\rule[-26mm]{0mm}{52mm}}
\includegraphics[width=7cm]{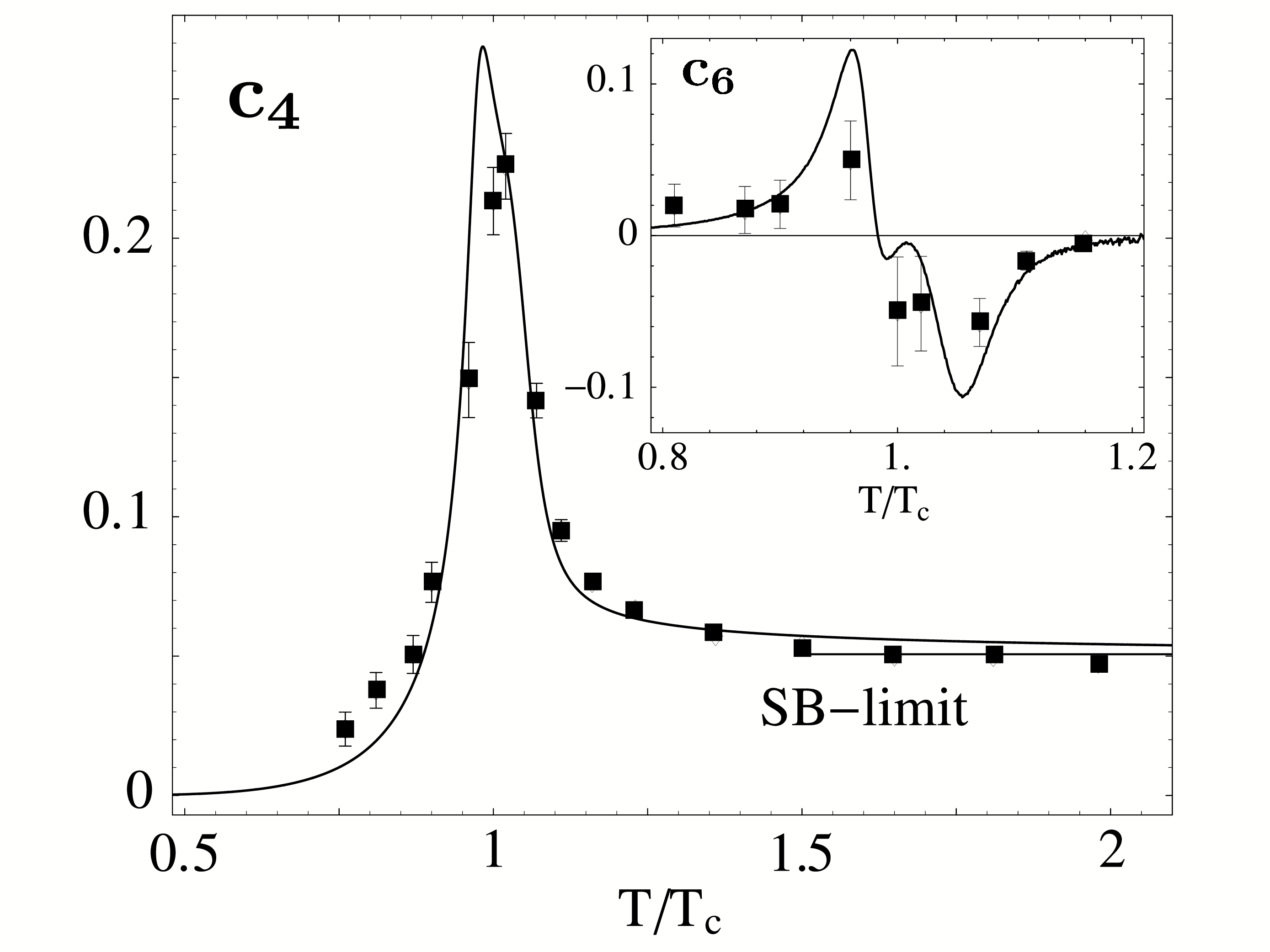}
\caption{Same as Fig.\ref{fig:5}, for the Taylor expansion coefficients $c_4(T)$ and $c_6(T)$. The expected high-temperture limits (SB) are also indicated.}
\label{fig:6}
\end{minipage}
\end{figure}

At non-zero quark chemical potential $\mu$, LQCD ``data" are deduced from Taylor expansions of thermodynamic quantities in powers of $\tilde{\mu}\equiv\mu/T$ around $\mu = 0$. For example, the pressure is expanded as
\begin{equation}
p(T,\mu)=T^4 \sum_{n=0}^{\infty}c_n\left(T\right)
\tilde{\mu}^n~~,~~~~~c_n\left(T\right)=\left.\frac{1}{n!}\frac{\partial^n (p/T^4)}
{\partial\tilde{\mu}^n}\right |_{\mu=0}.
\label{pressure}
\end{equation}
We can thus compare the lattice expansion coefficients $c_n(T)$ to those calculated in the PNJL model. Results for the first few coefficients $c_{2,4,6}$ are shown in Figs.\ref{fig:5},\ref{fig:6}. The comparison is evidently quite successful in view of the fact that no fine-tuning of parameters has been made, although one must keep in mind that the quark masses used in the LQCD simulations are not the small current quark masses used as input in the PNJL model.

%figure-----------------------------------------------------------------------------------------------------------------------------
%figure-----------------------------------------------------------------------------------------------------------------------------

\begin{figure}[tbh]
\centering
\includegraphics[width=7cm]{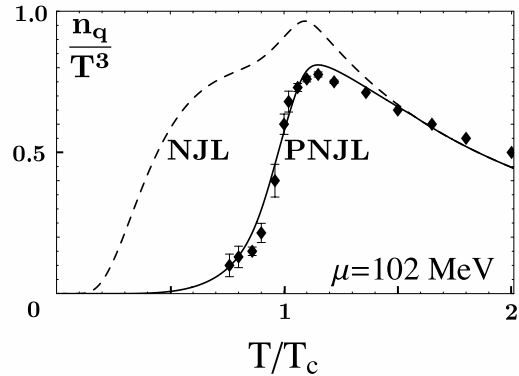}
\caption{Temperature dependence of the quark number density (\ref{density}) at $\mu \simeq.1$ GeV calculated in the PNJL model\cite{RTW06} (solid curve) and in the NJL model (with $\Phi \equiv 1$, dashed curve). Lattice QCD results\cite{Allton02} using the truncated Taylor expansion (\ref{density}) are shown for comparison.}
\label{fig:7}
\end{figure}

It is instructive to examine the quark number density,
\begin{equation}
n_q(T,\mu) = {\partial p(T,\mu)\over \partial\mu} = 2c_2\,\mu T^2 + 4c_4\,\mu^3 + 6c_6{\mu^5\over T^3} + . . .~~ ,
\label{density}
\end{equation}
and to compare the PNJL result, including Polyakov loop dynamics, with a calculation using the ``classic" NJL model in which quarks propagate freely without confinement constraints. Fig.\ref{fig:7} clearly demonstrates the important influence of the Polyakov loop: it suppresses the quarks as thermodynamically active quasiparticle degrees of freedom as the transition temperature $T_c$ is approached from above. Without this restriction, quarks would pile up in the ``forbidden" region below $T_c$.  

Further recent results concerning the quark number susceptibility, 
\begin{equation}
\chi_q(T,\mu) = {\partial^2 p(T,\mu)\over \partial\mu^2} = 2c_2\,T^2 + 12\,c_4\,\mu^2 + 30\,c_6{\mu^4\over T^2} + . . .~~ ,
\label{susc}
\end{equation}
have also been analysed\cite{RRW07b} using the PNJL model which, at this point, can help estimating uncertainties induced by truncation of the Taylor expansion in $\mu$.

%figure-----------------------------------------------------------------------------------------------------------------------------
%figure-----------------------------------------------------------------------------------------------------------------------------

\subsection{Phase diagram}

\begin{figure}[tbh]
\centering
\includegraphics[width=7cm]{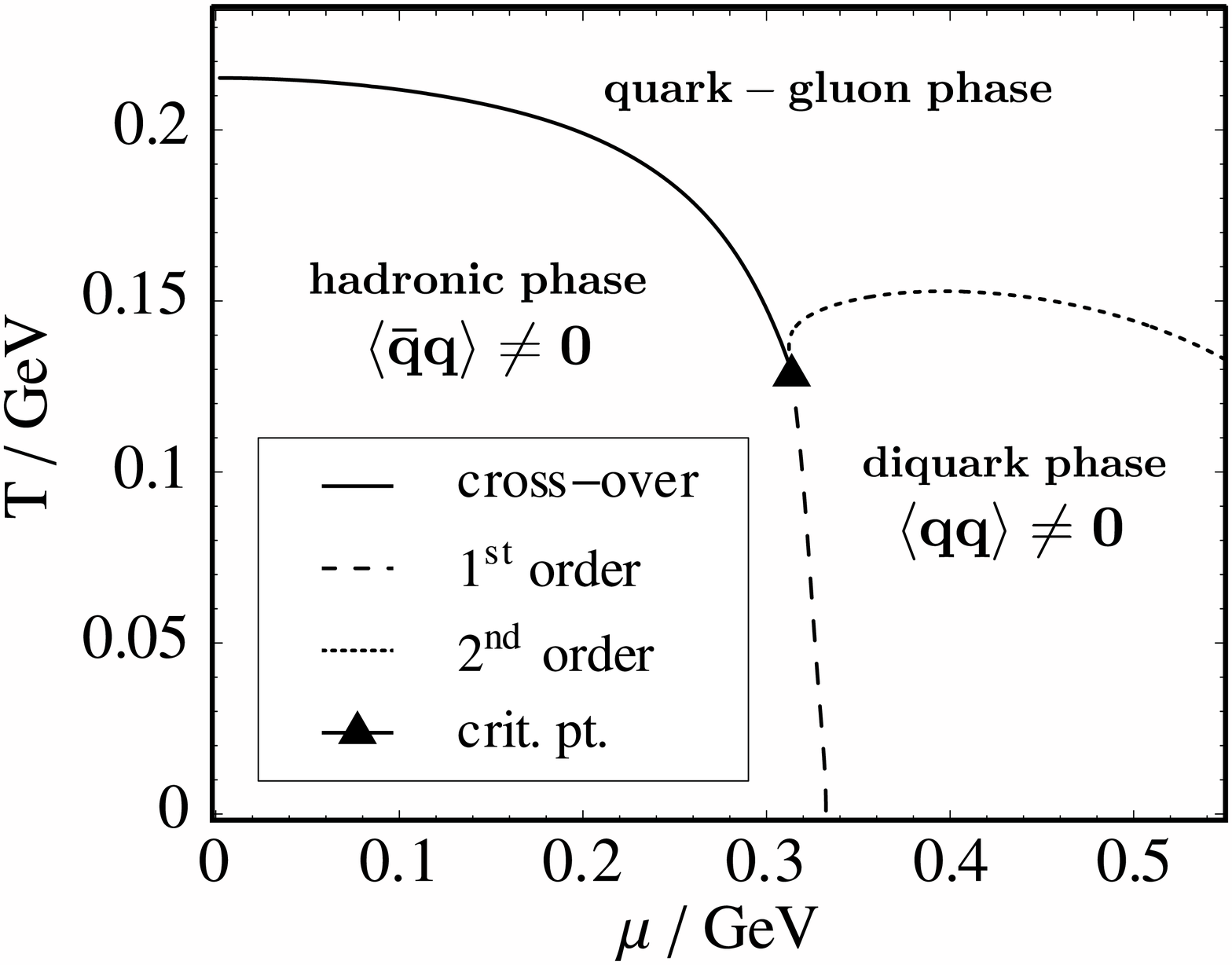}
\caption{Phase diagram calculated in the two-flavor PNJL model\cite{RRW07b}. The solid curve represents the chiral crossover region. Dashed curve: first order transition to superconducting high-density phase. Dotted curve: second order transition.} 
\label{fig:8}
\end{figure}

We finally turn to the phase diagram in the $(T,\mu)$ plane as derived\cite{RRW07b} from the two-flavor PNJL model. The result is shown in Fig.\ref{fig:8}. The solid line between the hadronic phase and the quark-gluon phase represents the chiral crossover transition. The dashed line marks a first order transition between the phase with spontaneously broken chiral symmetry and the (color) superconducting high-density phase in which diquarks accumulate to form a non-vanishing (Cooper pair) condensate. The transition between this phase and the quark-gluon sector (marked by the dotted line) is second order. Further inspection\cite{RRW07b} reveals that the critical endpoint shown in the figure is not a tricritical point; the three transition lines do not have a common junction. It turns out in fact that the location of this point is very sensitive to the input quark mass and to the presence or absence of Polyakov loop dynamics. For example, the temperature at which the gap $\Delta$ disappears is shifted upward by about 100 MeV when the Polyakov loop is active, as compared to the standard NJL model. A more detailed discussion can be found in ref.\cite{RRW07b}. 

%figure-----------------------------------------------------------------------------------------------------------------------------
%figure-----------------------------------------------------------------------------------------------------------------------------

\section{Conclusions and Outlook}

A quasiparticle approach (the PNJL model) encoding the two basic features that govern low-energy QCD, chiral symmetry and confinement, has been developed. It operates with ``order parameter" fields, the chiral condensate and the Polyakov loop, coupled through quarks as quasiparticles with dynamically generated masses. Despite its extreme simplicity, this approach turns out to be surprisingly successful in confrontations with $N_f = 2$ QCD thermodynamics on the lattice, at least for a temperature range up to about twice the critical temperature $T_c \sim 0.2$ GeV. 

Further developments now include extensions beyond mean field theory\cite{RRW07c}, and the generalisation to $2+1$ flavors with the additional effects of  Polyakov loop dynamics, in order to explore the rich variety of superconducting phases. Many more steps are still ahead, such as replacing the notorious NJL cutoff by a running coupling strength\cite{HRW07} and establishing contacts to the high temperature limit with incorporation of transverse gluons. Last not least, the hadronic matter phase with its composite (meson and baryon) degrees of freedom must be approached with an improved  effective field theory in order to match with a more realistic equation of state at high baryon density.

\section*{Acknowledgement}
W. W. thanks the Yukawa Institute for Theoretical Physics at Kyoto University and the organisers of YKIS2007 for their hospitality and for providing the stage and atmosphere of a most stimulating symposium.

\end{document}